\documentclass[superscriptaddress,showkeys,twocolumn,nofootinbib,longbibliography]{revtex4-1}

\usepackage{graphicx}
\usepackage{amsmath}
\usepackage{slashed}
\usepackage{feynmp}

\newcommand{\I}{\ensuremath{\mathrm{i}}}
\newcommand{\qm}[1]{``#1''} 
\renewcommand{\d}{\ensuremath{\mathrm{d}}}
\newcommand{\hc}{\ensuremath{\mathrm{h.c.}}}
\newcommand{\arctg}{\mathop{\rm arctg}\nolimits}

\usepackage{color}
\usepackage{ulem} 

\definecolor{myred}{rgb}{1,0,0}
\definecolor{mygreen}{rgb}{0,0.8,0.2}
\definecolor{myblue}{rgb}{0,0,1}

\definecolor{Ared}{rgb}{1,0.7,0}
\definecolor{Agreen}{rgb}{0.7,0.8,0.2}
\definecolor{Ablue}{rgb}{0,0.7,1}

\renewcommand{\emph}[1]{\textit{#1}}

\usepackage[nodayofweek]{datetime}
\newdateformat{mydate}{\twodigit{\THEDAY}{ }\shortmonthname[\THEMONTH], \THEYEAR}

\begin{document}

\title{Electroweak $SU(2)_L \times U(1)_Y$ model\\ with strong spontaneously fermion-mass-generating gauge dynamics}

\author{Petr Bene\v{s}}
\affiliation{Institute of Experimental and Applied Physics, Czech Technical University in Prague,\\Husova 240/5, 110~00 Prague 1, Czech Republic}

\author{Ji\v{r}\'{\i} Ho\v{s}ek}
\email{hosek@ujf.cas.cz}
\affiliation{Institute of Experimental and Applied Physics, Czech Technical University in Prague,\\Husova 240/5, 110~00 Prague 1, Czech Republic}
\affiliation{Department of Theoretical Physics, Nuclear Physics Institute, Czech Academy of Sciences, 250~68
\v{R}e\v{z} (Prague), Czech Republic}

\author{Adam Smetana}
\affiliation{Institute of Experimental and Applied Physics, Czech Technical University in Prague,\\Husova 240/5, 110~00 Prague 1, Czech Republic}

\begin{abstract}
Higgs sector of the Standard model (SM) is replaced by quantum flavor dynamics (QFD), the gauged flavor $SU(3)_f$ symmetry with scale $\Lambda$. Anomaly freedom requires addition of three $\nu_R$. The approximate QFD Schwinger-Dyson equation for the infrared fermion self-energies $\Sigma_f(p^2)$ has the \emph{spontaneous-chiral-symmetry-breaking solutions} ideal for seesaw: (1) $\Sigma_f(p^2)=M_{fR}^2/p$ where three Majorana masses $M_{fR}$ of $\nu_{fR}$ are of order $\Lambda$. (2) $\Sigma_f(p^2)=m_f^2/p$ where three Dirac masses $m_f=m_{(0)}1+m_{(3)}\lambda_3+m_{(8)}\lambda_8$ of SM fermions are {\it exponentially suppressed w.r.t. $\Lambda$}, and {\it degenerate for all SM fermions in $f$}.
(1) $M_{fR}$ break $SU(3)_f$ symmetry completely; $m_{(3)},m_{(8)}$ superimpose the tiny breaking to $U(1) \times U(1)$.
All flavor gluons thus acquire self-consistently the masses $\sim \Lambda$.
(2) All $m_f$ break the electroweak $SU(2)_L \times U(1)_Y$ to $U(1)_{em}$.
Symmetry partners of the composite Nambu-Goldstone bosons are the genuine Higgs particles: (1) Three
$\nu_{R}$-composed Higgses $\chi_i$ with masses $\sim \Lambda$. (2) Two new SM-fermion-composed Higgses $h_3, h_8$ with masses
$\sim m_{(3)}, m_{(8)}$, respectively. (3) The SM-like SM-fermion-composed Higgs $h$ with mass $\sim m_{(0)}$, the effective Fermi scale.
$\Sigma_f(p^2)$-dependent vertices in the electroweak Ward-Takahashi identities imply:
The axial-vector ones give rise to the $W$ and $Z$ masses at Fermi scale.
The polar-vector ones give rise to the fermion mass splitting in $f$.
At the present exploratory stage the splitting comes out unrealistic.
\end{abstract}


\maketitle

\section{Introduction}
Principle of spontaneous symmetry breaking, ingeniously realized in
the SM of electroweak interactions \cite{sm} in the form of the
Higgs mechanism \cite{higgs} turned out to be extremely successful.
First, three `would-be' Nambu-Goldstone (NG) bosons, pre-prepared in
the spontaneously condensing elementary scalar electroweak complex
doublet Higgs field give rise to the longitudinal polarization
states of massive $W^{\pm}$ and $Z$ bosons. Second, the condensate
of the Higgs field, which defines the Fermi (electroweak) scale
gives rise to the fermion masses in terms of {\it theoretically
arbitrary} (infinitely renormalized) Yukawa coupling constants.
Third, the observable remnant of the Higgs mechanism is the real
scalar above the condensate which remains in the Higgs field. It
describes the massive spinless Higgs boson $H$  with uniquely
prescribed interactions. {\bf In 2012 such a boson was discovered at
the CERN LHC} \cite{cern}. The resulting picture, confirmed
experimentally with steadily increasing accuracy is theoretically
consistent all the way up to the Planck scale.

On this way, however, the Higgs mechanism as a source of particle
masses is not complete; it is blind to three important facts:
First, the neutrinos are
massive. Second, there is the dark matter. Third, if the CERN Higgs
were indeed the Higgs boson of the Standard model the lepton and
quark masses would be the phenomenological, theoretically arbitrary
parameters forever. The spectra of quantum systems are, however, as
a rule, calculable.

Fortunately, principles are more general than their particular
realizations. Already the masters of the Higgs mechanism, knowing
that the same principle of spontaneously broken symmetry is realized
also in superconductors, pointed out (P. Higgs in \cite{higgs}, F.
Englert and R. Brout in \cite{eb} that the Higgs boson could be a
composite fermion-anti-fermion particle. Literature on this subject
is vast \cite{cpnsh}. The very description of bound states in
relativistic quantum field theory requires strong coupling and is
therefore exceedingly difficult to handle. Consequently, no model of
this sort which would be comparably convincing as the weakly coupled
Higgs mechanism is available. Our suggestion presented here is no
exception.

We refer with admiration to the papers of Heinz Pagels and co-workers
\cite{pagels-stokar}, \cite{carter-pagels}, \cite{pagels}: These papers
analyze some {\it consequences} of the dynamical Higgs mechanism which
they call quantum flavor dynamics (QFD) \cite{pagels-stokar}.
The main points of their QFD, dealing with SM fermions and a new gauge dynamics,
without specifying the Lagrangian, are common with the approach suggested here:
First, the dynamically generated fermion masses are finite and calculable
\cite{pagels-stokar}. Second, their dynamical, spontaneous
generation implies the masses of $W$ and $Z$ bosons \cite{carter-pagels}.
Third, there is a composite Higgs particle \cite{carter-pagels}.
We find the name QFD very appropriate and take the liberty of using it for our
gauge quantum flavor dynamics (QFD) defined by gauging properly the flavor $SU(3)$ symmetry index.

It is obvious that if the SM fermion masses are generated (somehow)
dynamically the electroweak gauge boson masses come out as the
necessary consequence of the existence theorem of Goldstone. The
composite Higgs is then the plausible symmetry partner of the
composite `would-be' NG bosons. Its existence is not, however,
guaranteed by an existence theorem. This idea was elaborated
heuristically in several infamous papers \cite{infam} as well as in
several famous ones \cite{fam}. Majority of them of course quote the
basic Ref.\cite{carter-pagels}.

In the present model the SM fermion self-energies $\Sigma_f(p^2)$
generated by the strong-coupling QFD are by symmetry the same in
each flavor $f$. This does not correspond to reality at all.
Fortunately, there are the electroweak interactions. The necessary
appearance of $\Sigma_f(p^2)$ in the {\bf electroweak Ward-Takahashi
identities} seems to supplement naturally the QFD.

The paper is organized as follows: In Sect.II we define the QFD and
argue that it generates spontaneously the fixed fermion mass
pattern. In Sect.III we describe the consequences of this pattern
dictated by the powerful Goldstone theorem supplemented with the
Nambu-Jona-Lasinio-like assumption \cite{njl} on the existence of
the NG symmetry partners. In Sect.IV we discuss the effects of
$\Sigma_f(p^2)$ in the electroweak WT identities: The axial-vector
terms with NG poles have the robust effect of generating the $W$ and $Z$ boson masses
with observed ratio given by the Weinberg angle.
The polar vector terms which we suggest for computing the fermion mass splitting
in $f$ are strongly model-dependent. The present computation has therefore only the illustrative character.
Sect.V contains our brief general conclusions.

\section{Properties of QFD}

Gauging the flavor or family, or horizontal, or generation symmetry
is so natural that it cannot be new\cite{gauging-flavor}. Being
completely and badly broken it has to be broken spontaneously. The
safe way is to use the weakly coupled Higgs sector. There is the
arbitrariness in choosing the gauge group, the assignment of the
chiral fermion fields to the representations of the gauge group, and
the choice of the Higgs fields.

Here we consider the real world with three SM fermion chiral families, i.e., the gauge flavor dynamics is $SU(3)_f$. We put all chiral SM
fermion multiplets into triplets
of the flavor group. The sacred requirement of anomaly freedom then
uniquely implies adding one triplet of the right-handed neutrino
fields $\nu^f_{R}$. This is
the starting point of the model of Tsutomu Yanagida \cite{yanagida}.
Its fermion and flavor gluon Lagrangian has the form
\begin{eqnarray}
{\cal L}_f &=& 
- \frac{1}{4} F_{a\mu\nu} F_a^{\mu\nu} 
+ \bar q_L \I\slashed{D} q_L 
+ \bar u_R \I\slashed{D} u_R 
+ \bar d_R \I\slashed{D} d_R
\nonumber \\ && {} 
\phantom{- \frac{1}{4} F_{a\mu\nu} F_a^{\mu\nu}}
+ \bar l_L \I\slashed{D} l_L 
+ \bar e_R \I\slashed{D} e_R 
+ \bar \nu_{R} \I\slashed{D} \nu_{R}
\,.
\hspace{3mm}
\end{eqnarray}

Yanagida himself \cite{yanagida} describes the observed broken gauge
flavor and electroweak gauge symmetries with two vastly different
mass scales put by hands by an extended sector of weakly interacting
Higgs fields. They are uniquely defined by the available
gauge-invariant Yukawa couplings. Be it as it may, this slightly
extended Standard model is an elegant quantum field theory
realization of the seesaw \cite{seesaw}, and also the basis of
understanding the baryogenesis via leptogenesis \cite{leptogenesis}.

In conclusion of his paper Yanagida suggests that the model is \qm{a possible
candidate for the spontaneous mass generation by dynamical symmetry breaking},
having in mind the famous model of Nambu and Jona-Lasinio \cite{njl}.
The present model wants to be a honest attempt in realizing this suggestion:
We pretend to obtain two vastly different mass scales spontaneously as
two distinct solutions
in two different channels of one matrix QFD Schwinger-Dyson equation
(gap equation) for the chirality-changing fermion self-energies $\Sigma(p^2)$.

The field tensor $F$ describes the kinetic term of eight flavor gluons $C$
and their self-interactions, the covariant derivatives $D$ describe
their interactions with chiral, both right- and left-handed, fermions. The $q_L$ and $l_L$ are the
quark and lepton electroweak doublets, respectively, $u_R, d_R, e_R,
\nu_R$ are the electroweak singlets. Their weak hypercharges are uniquely fixed by the corresponding
electric charges.

The gauge $SU(3)_f$ interaction, characterized by one dimensionless coupling
constant $h$, is {\it asymptotically free} at high momenta and hence
{\it strongly interacting in the infrared}. This means that by dimensional
transmutation the dimensionless $h$ can turn into the theoretically
arbitrary mass scale $\Lambda$.

The Lagrangian ${\cal L}_f$ is always considered together with the standard
electroweak gauge $SU(2)_L
\times U(1)_Y$ forces known to remain weakly coupled all the way up to
the Planck scale: In the first stage, when considering the effects of the
strong-coupling QFD we take the electroweak interactions only as a weak
external perturbation;
basically, QFD should only know what are the electric charges of the fermions present.
In the second stage we take the electroweak interactions into account perturbatively.
We are interested in the effects of  $\Sigma_f(p^2)$ which appear in the electroweak
Ward-Takahashi identities. The QCD does not influence the phenomena described here.

Like in the QCD Lagrangian, also in ${\cal L}_f$ both the left- and the
right-handed fermion fields interact with the octet
of the corresponding masless gauge bosons as $SU(3)$ triplets.
This similarity is highly suspicious.
On the first sight it seems that what we suggest contradicts the Vafa-Witten
no-go theorem \cite{vafa-witten}.
Since in QCD we trust we are obliged to provide a truly good reason why in
the infrared the suggested QFD should self-break, generating spontaneously
the masses of its elementary excitations (leptons, quarks and flavor gluons),
whereas the QCD confines.

We believe there is such a good reason: While the QCD deals with the
electrically charged quarks, the QFD deals also with the
electrically neutral sterile right-handed neutrinos {\it which can be the
massive Majorana particles}.
It is utmost important that their hard mass term
\begin{eqnarray}
\label{majorana}
{\cal L}_{\mathrm{Majorana}} &=& - \frac{1}{2} \bar \nu_{R}M_{R}(\nu_{R})^{{\cal C}} + \hc
\end{eqnarray}
{\it unlike the Dirac mass term}, is strictly prohibited by the $SU(3)_f$ gauge symmetry:
It transforms as $3^{*} \times 3^{*} = 3 + 6^{*}$
which does not contain unity. It can, however, be generated dynamically
\`{a} la Nambu-Jona-Lasinio \cite{njl} provided this option is energetically
favorable (which we assume). It is easy to see that the relevant part of the Lagrangian
\begin{eqnarray}
{\cal L}_{\rm int} &=& 
\frac{1}{2}h
\Big(\bar
\nu_{R}\gamma_{\mu}\tfrac{1}{2}\lambda_a\nu_{R}+ \overline
{(\nu_{R})^{{\cal
C}}}\gamma_{\mu}[-\tfrac{1}{2}\lambda_a^T](\nu_{R})^{{\cal
C}}
\Big)C^{\mu}_a
\hspace{8mm}
\end{eqnarray}
in contrast with the vector-like QCD Lagrangian, is effectively
chiral: The charge conjugate neutrino field $(\nu_R)^{{\cal C}}= C
(\bar \nu_R)^T$ is of course a {\it left-handed field},
$(\nu_R)^{{\cal C}}=(\nu^{\cal C})_L$. Unlike the other left-handed
fields in the Lagrangian ${\cal L}_f$ it transforms, however, as the
{\it antitriplet} of $SU(3)_f$: $T_a(L)=-\tfrac{1}{2}\lambda_a^{T}$.
An answer to the mandatory question why the Majorana masses of the
left-handed neutrinos are not dynamically generated is suggested in
\cite{hosek-seesaw}.

The strategy is in principle crystal-clear: First, in the Lagrangian dealing
only with chiral fermions and with
gauge fields {\it all hard fermion mass terms} are strictly prohibited by the underlying
gauge chiral $SU(3)_f \times SU(2)_L \times U(1)_Y$ invariance \cite{argument}.
Second, if the appropriate chirality-changing fermion self-energies $\Sigma(p^2)$
are spontaneously generated
by the QFD strongly coupled in the infrared, they fix the pattern of
spontaneously broken symmetries uniquely.
The powerful Goldstone theorem then yields a number of strong conclusions.
We supplement the Goldstone theorem with a plausible assumption
of the existence of the genuine symmetry partners of the composite `would-be'
NG bosons. It is gratifying that the generic properties of the SM
are reproduced. Harsh support or invalidation of our suggestion can apparently
be given only by the non-perturbative lattice computations.
For gauge theories with chiral fermions they are, however, at present not
available \cite{lattice-chiral}.\\

Our primary aim therefore is to find the non-perturbative solutions of the
Schwinger-Dyson (SD) equation for the chiral-symmetry-breaking $\Sigma(p^2)$ in
the full fermion propagators $S^{-1}(p)=\slashed p - \Sigma(p)$
for all fermions of the $SU(3)_f \times SU(2)_L \times U(1)_Y$ gauge-invariant Lagrangian.
The matrix SD equation of QFD (in Landau gauge) is \cite{pagels}
\begin{widetext}
\begin{eqnarray}
\Sigma(p) &=& 3\int \frac{\d^4k}{(2\pi)^4}\frac{\bar
h^2_{ab}\big((p-k)^2\big)}{(p-k)^2}T_a(R)\Sigma(k)\Big[k^2+\Sigma^{+}(k)\Sigma(k)\Big]^{-1}T_b(L)
\,.
\label{Sigma}
\end{eqnarray}
\end{widetext}
According to the NJL self-consistent reasoning \cite{njl} we first {\it assume}
that the gauge flavor $SU(3)_f$ is completely
self-broken, and subsequently {\it find} the corresponding symmetry-breaking solutions.

It is important to realize that the SD equation is universal: (i) The Majorana
mass is the R-L bridge between
$\nu_R$ which transforms as a flavor triplet, and between the left-handed
$(\nu_R)^{{\cal C}}$
which transforms as flavor anti-triplet ($3^{*} \times 3^{*} = 3 + 6^{*}$).
The corresponding $\Sigma(p^2)$ which gives rise
to three Majorana masses $M_{fR}$ is the complex $3 \times 3$ matrix, the
symmetric sextet by Pauli principle,
and $T_a(R)=\tfrac{1}{2}\lambda_a$,   $T_a(L)=-\tfrac{1}{2}\lambda_a^T$.
(ii) The Dirac mass is the R-L bridge between the right- and the left-handed fermion fields
both transforming as flavor triplets ($3^{*} \times 3 = 1 + 8$):
The corresponding $\Sigma(p^2)$ which gives rise to three Dirac masses $m_f$
is a general complex $3 \times 3$ matrix, and $T_a(R)=T_a(L)=\tfrac{1}{2}\lambda_a$.

{\it Because there is nothing in QFD which would distinguish between
the neutrino, the charged lepton, the charge $Q=2/3$ quark, and the
charge $Q=-1/3$ quark in given family {\bf the masses $m_f$ must come out
degenerate for all fermion species in family $f$}. Difference between
the Majorana and Dirac mass matrices turns out, however, substantial.}

We have demonstrated elsewhere \cite{hosek} that in a separable
approximation for the kernel of the SD equation (\ref{Sigma}) there
are explicit solutions for $\Sigma(p^2)$ with the following
properties:

(1) There are three Euclidean  Majorana self-energies $\Sigma_{f}(p^2)=M_{fR}^2/p$
where the three Majorana masses $M_{fR}$ are
\begin{equation}
\boxed{M_{fR} \sim \Lambda}
\end{equation}
(2) There are three Euclidean Dirac self-energies $\Sigma_{f}(p^2)=m_f^2/p$
where the three Dirac masses $m_f$ degenerate for $\nu_f, e_f, u_f, d_f$
in family $f$ are exponentially small with respect to $\Lambda$:
\begin{equation}
\boxed{
m_f = \Lambda \, \exp(-1/4\alpha_{f})
}
\label{mD}
\end{equation}
Here $\alpha_f$ are three dimensionless effective coupling constants of separable approximation.\\

We note that the paradigm-changing super-conducting gap of BCS
\cite{bcs} was also the result of a weird separable approximation:
In the gap equation it simply ignored the vast majority of
interactions of electrons in superconductors. Only many years later
this issue was clarified by Polchinski \cite{polchinski}.

Another important note is this \cite{hosek}: The obtained functional
behavior of $\Sigma(p^2) \sim 1/p$ is apparently good only in the
infrared (IR), i.e., at low momenta where the momentum-dependent
coupling is large. There it defines the fermion masses
\cite{mannheim}. In solving the SD equation in separable
approximation the integration over momenta runs only up to
$\Lambda$. At high momenta, i.e., in the ultraviolet (UV) the
behavior of $\Sigma(p^2)$ is dictated by asymptotic freedom of QFD
\cite{pagels}: $\Sigma(p^2) \sim 1/p^2$. In our approximation the
QFD is not asymptotically,  but entirely free. As the fermion mass
is a low-momentum phenomenon
the high-momentum regime is not essential for our purposes.\\

How many free parameters are {\it ultimately} necessary for
computing the fermion masses in QFD? The strong non-Abelian
$SU(3)_f$ dynamics is characterized by one theoretically arbitrary
parameter, the scale $\Lambda$. Consequently, {\it if our basic
strong assumption of the complete self-breaking is warranted} both
the Majorana masses $M_{fR}$ and the Dirac masses $m_f$ should
ultimately be the {\it calculable} multiples of $\Lambda$
\cite{pagels-stokar}. In any case, in the sterile neutrino sector
the sextet is characterized by three vacuum expectation values
\cite{bhs} as is the singlet plus octet in the SM fermion sector.
Clearly, the flavor gluon dynamics uniquely relates them.

Our belief here is entirely analogous to the belief in understanding
the hadron mass spectrum of the confining QCD in the chiral limit:
With one theoretically arbitrary scale $\Lambda_{QCD}$ there are the
massless NG pions, whereas the masses of all other hadrons of the
first family ($m_u, m_d \ll \Lambda_{QCD}$) are ultimately the
calculable multiples of $\Lambda_{QCD}$ (so far only approximately
by a computer). The case of QFD is even more complex because besides
the masses of its elementary excitations (leptons, quarks and flavor
gluons) there are also the masses of its expected unconfined but
strongly coupled collective excitations or bound states. Putting the
effectively chiral QFD on the lattice will be hard \cite{lattice-chiral},
because the hard Majorana masses of sterile neutrinos are strictly
prohibited by symmetry: $3 \times 3 = 3^{*} + 6$ does not contain
unity.

\section{Goldstone theorem implies}

{\it We will assume in the following that the fermion mass pattern $M_{fR} \gg m_f$
obtained in a crude approximation is the generic property of QFD
at strong coupling}. It then follows that it fixes the spontaneous symmetry-breaking pattern
of the underlying gauge $SU(3)_f \times SU(2)_L \times U(1)_Y$ symmetry down to $U(1)_{em}$
uniquely, and the Goldstone theorem implies
several strong reliable conclusions. The Goldstone theorem is supplemented with a
plausible assumption
of the existence of the genuine symmetry partners of the composite `would-be' NG bosons.
For transparency we rewrite $m_f$ as ($\lambda_0=\surd \tfrac{2}{3} {\bf 1}$)
\begin{equation}
m_f \equiv 
  m_{(0)} \lambda_0 
+ m_{(3)} \tfrac{1}{2}\lambda_3 
+ m_{(8)} \tfrac{1}{2}\lambda_8
\,,
\end{equation}
where
\begin{subequations}
\begin{eqnarray}
m_{(0)} &=& \tfrac{1}{\surd 6}(m_1 + m_2 + m_3) \,,\\
m_{(3)} &=& m_1 - m_2 \,,\\
m_{(8)} &=& \tfrac{1}{\surd 3}(m_1 + m_2 - 2m_3) \,.
\end{eqnarray}
\end{subequations}\\

I. (1) The Majorana masses $M_{fR}$ break down the gauge symmetry $SU(3)_f$ spontaneously
and completely \cite{yanagida}, \cite{bhs}, \cite{hosek}. Consequently, {\bf eight
`would-be' NG bosons composed of sterile neutrinos give rise to masses $m_{iC}$ of all
flavor gluons $C_i^{\mu}$ of order $\Lambda$} \cite{migdal-polyakov},
\cite{jackiw-johnson}, \cite{cornwall-norton}.
In the $SU(3)_f$ WT identities for the sterile neutrino sector \cite{hosek} the
NG bosons are convincingly identified as massless poles together with their quantum numbers:
some are scalars, some are pseudo-scalars.\\

(2) Eight `would-be' NG bosons and one genuine pseudo NG boson resulting
from spontaneous breakdown of global
anomalous $U(1)$ symmetry of the right-handed neutrino sector belong to
the complex symmetric composite sextet \cite{hosek}
\begin{equation}
\Phi_{\{fg\}} = \frac{1}{\Lambda^2} \big(\bar \nu_{fR}(\nu_{gR})^{{\cal C}}\big)
\end{equation}
Consequently, as a remnant of symmetry ($8 + 1 + 3 = 12$), {\bf there should exist three genuine
Higgs-like composite bosons $\chi_i$ with masses of order  $\Lambda$} \cite{hosek}.\\

II. (1) The Dirac mass $\bar\psi_{fR}m_{(0)}\psi_{fL}+ \hc$ is $SU(3)_f$ invariant, i.e.,
the QFD symmetry in isolation would allow the hard Dirac fermion mass common to all fermions
  of three families. Such a term breaks, however, the electroweak $SU(2)_L \times U(1)_Y$
  chiral symmetry spontaneously down to $U(1)_{em}$ even if that is considered as a weak
  external perturbation. Consequently, {\bf three multi-component NG bosons
  composed of all electroweakly interacting
leptons and quarks are dynamically generated}. Only in the second stage, when the
gauge electroweak interactions are switched on these NG bosons become `would-be',
giving incoherently rise to masses $m_W$ and $m_Z$ of $W$ and $Z$ bosons, respectively,
in terms of $\sum m_f$, the induced Fermi (electroweak) scale.
In the $SU(2)_L \times U(1)_Y$ WT identities for the SM fermions \cite{hosek},
see Eqn.(\ref{gammaW}), (\ref{gammaZ}) the NG bosons are convincingly identified as massless poles together
with their quantum numbers:
as the fermion masses in families are degenerate they are the pure pseudo-scalars.
For the same reason the canonical Weinberg relation $m_W/m_Z = \cos \theta_W$ is exact.
We note a difference between the composite `would-be' NG bosons of spontaneously
broken symmetries $SU(3)_f$ and $SU(2)_L \times U(1)_Y$:
In both cases they are composed by the strong-coupling QFD, but in the former
case they are always `would-be'.
\\

(2) Three `would-be' NG bosons belong to the complex multi-component composite doublet
(index $a$ in the following formula)
\begin{eqnarray}
\phi^a &=& \frac{1}{\Lambda^2} \Big[ (\bar e_{fR} l^{af}_{L}) + (\bar d_{fR} q^{af}_{L}) 
\nonumber \\ && {}
\hspace{5mm}
+ \big(\overline {(\nu_{fR})^{{\cal C}}} (l^{af}_{L})^{{\cal C}}\big) + \big(\overline {(u_{fR})^{{\cal C}}} (q^{af}_{L})^{{\cal C}}\big)\Big]
\,.
\hspace{10mm}
\end{eqnarray}
For the $SU(2)$ spinors $\phi$ the definition of charge conjugation includes the
definition $\phi^{{\cal C}}=\I\tau_2 \phi^{*}$.
Consequently, as a remnant of symmetry (as in the Standard model)
($3 + 1 = 4$), {\bf there is the mandatory retro-diction of one genuine multi-component
composite SM-like Higgs boson $h$ with mass at the induced electroweak scale}.\\

III. (1) The Dirac masses $m_{(3)}$ and $m_{(8)}$ break down spontaneously the
flavor $SU(3)_f$ gauge symmetry in the SM sector down to $U(1) \times U(1)$ \cite{higgs, hosek},
and the electroweak $SU(2)_L \times U(1)_Y$ chiral symmetry spontaneously down to $U(1)_{em}$.
Consequently, {\bf six multi-component `would-be' NG bosons composed of the electroweakly
interacting leptons and quarks contribute a tiny amounts
to the huge masses of six flavor gluons} \cite{higgs}, and three multi-component
`would-be' NG bosons, also composed of the electroweakly interacting leptons and quarks
provide extra contributions to  $m_W$ and $m_Z$. The canonical Weinberg relation between them remains intact. \\

(2) All multi-component `would-be' NG bosons discussed above are contained in the composite operator
\begin{eqnarray}
\phi^a_i &=& \frac{1}{\Lambda^2} \Big[ (\bar e_{R}\tfrac{1}{2}\lambda_i l^{a}_{L}) + (\bar d_{R}\tfrac{1}{2}\lambda_i q^{a}_{L})
\nonumber \\ && {}
\hspace{5mm}
+ \big(\overline {(\nu_{fR})^{{\cal C}}}\tfrac{1}{2}\lambda_i (l^{a}_{L})^{{\cal C}}\big) + \big(\overline {(u_{fR})^{{\cal C}}}\tfrac{1}{2}\lambda_i (q^{a}_{L})^{{\cal C}}\big) \Big]
\,.
\hspace{8mm}
\end{eqnarray}

Important is to consider those fermion bilinear combinations of the
chiral fermion fields $\psi_{L,R}$ with the same electric charge, i.e., those which are responsible for the fermion masses $m_{(3)}$
and $m_{(8)}$, i.e., $\bar \psi_R \lambda_{3,8} \psi_L$. They belong
to the $SU(3)$ octet. To have it real we have in mind that its
redundant component becomes the `would-be' NG boson of otherwise
neglected $SU(2)_L \times U(1)_Y$. Consequently, as a remnant of
symmetry ($6 + 2 = 8$) {\bf there should exist two additional
multi-component composite Higgs-like bosons $h_3$ and $h_8$ with
characteristic Yukawa couplings and masses at the electroweak
scale}. It is rather remarkable that namely such a possibility was
explicitly mentioned as an example of the non-Abelian Higgs
mechanism by Peter Higgs in his seminal `Abelian' paper
\cite{higgs}.\\

IV. There are interesting phenomena associated with spontaneous breakdown of global chiral
Abelian symmetries of the model \cite{hosek}.
The anomalous ones result in observable new axion-like particles, the anomaly-free one can be gauged, resulting in
new massive  $Z'$ gauge boson.

\section{Electroweak WT identities imply}

So far we have analyzed the consequences of spontaneous emergence of
$\Sigma_f(p^2)$ generated by the strongly coupled QFD. The
chiral-symmetry-breaking $\Sigma_f(p^2)$s manifest operationally by
new terms also in the {\it electroweak interactions by virtue of the Ward-Takahashi identities}. The WT identities
have to remain valid regardless of whether the underlying symmetry
were broken spontaneously or not. Consequently, the new $\Sigma_f(p^2)$-dependent symmetry-breaking terms
in them should have the
important and theoretically reliable implications. The EW WT identitites are (we temporally
omit the flavor index $f$) \cite{pagels-stokar}, \cite{delbourgo},
\cite{benes-hosek}
\begin{widetext}
\begin{eqnarray}
(p'-p)_{\mu}\Gamma_A^{\mu}(p',p) &=& eQ_i \big[S^{-1}(p') - S^{-1}(p)\big] \,,
\\
(p'-p)_{\mu}\Gamma_W^{\mu}(p',p) &=& \frac{e}{2 \surd 2 \sin\theta_W}\Big\{ \big[S^{-1}(p')T^{+} - T^{+}S^{-1}(p)\big] - \big[S^{-1}(p')T^{+}\gamma_5 + T^{+}\gamma_5 S^{-1}(p)\big] \Big\}
\,,
\\
(p'-p)_{\mu}\Gamma_Z^{\mu}(p',p) &=& \frac{e}{\sin 2 \theta_W} \Big\{ \big[S^{-1}(p')\big(T_{3L}^i - 2 Q_i  \sin^2 \theta_W\big)-\big(T_{3L}^i - 2 Q_i \sin^2 \theta_W\big)S^{-1}(p)\big]
\nonumber \\ && {}
\hspace{11mm}
- \big[S^{-1}(p')T_{3L}^i\gamma_5 + T_{3L}^i\gamma_5 S^{-1}(p)\big] \Big\}
\,.
\end{eqnarray}
\end{widetext}
The index $i$, ($i=\nu, l, u, d$) distinguishes different SM fermion
species. The proper vertices themselves which satisfy the WT
identities and have no unwanted kinematic singularities have the
form
\begin{widetext}
\begin{eqnarray}
\Gamma_A^{\mu}(p',p) &=& eQ_i \big[\gamma^{\mu} - (p' + p)^{\mu}\Sigma'(p',p)\big] \,,
\\
\Gamma_W^{+\mu}(p',p) &=& \frac{e}{2 \surd 2 \sin\theta_W} \Big\{ \big[\gamma^{\mu} - (p' + p)^{\mu}\Sigma'(p',p)\big]T^{+} - \big[\gamma^{\mu}\gamma_5 T^{+}-\tfrac{(p'-p)^\mu}{(p'-p)^2}\big(\Sigma(p')+\Sigma(p)\big)\gamma_5 T^{+}\big] \Big\} \,,
\label{gammaW}
\\
\Gamma_Z^{\mu}(p',p) &=& \frac{e}{\sin 2 \theta_W} \Big\{ \big[\gamma^{\mu} - (p' + p)^{\mu}\Sigma'(p',p)\big](T_{3L}^i - 2 Q_i  \sin^2 \theta_W)-\big[\gamma^{\mu}\gamma_5-\tfrac{(p'-p)^{\mu}}{(p'-p)^2}\big(\Sigma(p')+\Sigma(p)\big)\gamma_5\big] T_{3L}^i \Big\} \,.
\hspace{10mm}
\label{gammaZ}
\end{eqnarray}
\end{widetext}
The 'derivative' is defined as
\begin{eqnarray}
\Sigma'(p',p) &\equiv& \frac{\Sigma(p') - \Sigma(p)}{p'^2 - p^2} \,.
\end{eqnarray}
It in fact points to the difference between the
$\Sigma(p^2)$-dependent terms in the axial-vector and polar-vector
vertices. In the former ones they mark the famous massless
`would-be' Nambu-Goldstone poles and persist even for constant
$\Sigma$s, i.e., for hard fermion masses. This is how the NG pole
manifests in the NJL model. This robust effect is supported by the existence theorem.

In polar-vector vertices the appearance of the chiral-symmetry
breaking $\Sigma$s is much more subtle. They appear only
provided they are momentum-dependent. Consequently, their consequences
depend crucially upon their functional form and in evaluating their
credibility we should be very humble.

\subsection{Masses of $W$ and $Z$ bosons}
How the composite `would-be' NG bosons become the longitudinal
polarization states of massive $W$ and $Z$ bosons is well known. In
the present model the mechanism was described in detail in
\cite{hosek} and here we merely quote the result. For the Euclidean
$\Sigma_f(p^2)= m_f^2/p$ the new axial-vector vertices contribute to
the gauge-boson polarization tensor loop and the electroweak
gauge-boson masses are given by the famous Pagels-Stokar formula in
the form of the sum rules. The technique is identical to that used
in technicolor \cite{technicolor}.
\begin{equation}
\boxed{
m_W^2=\tfrac{1}{4}g^2\tfrac{5}{4\pi} \sum_f m_f^2=\tfrac{5\alpha}{4\sin^2 \theta_W}\sum_f m_f^2
}
\label{mW}
\end{equation}

\begin{equation}
\boxed{
m_Z^2=\tfrac{1}{4}(g^2 + g'^2)\tfrac{5}{4\pi} \sum_f m_f^2=\tfrac{5\alpha}{4\sin^2 \theta_W \cos^2 \theta_W}\sum_f m_f^2
}
\label{mZ}
\end{equation}
The Weinberg relation $m_W=m_Z \cos \theta_W$
is the consequence of the degeneracy of the fermion masses
in electroweak doublets. It is due to the fact that all chiral fermion fields are in flavor triplets.
If some strong dynamics would produce spontaneously the fermion masses with large splitting,
the relation would be badly broken \cite{carter-pagels}.

\subsection{SM-fermion mass splitting}

Inspired by prescient Heinz Pagels who attempted to compute the
$u-d$ quark mass difference due to the electromagnetic interaction
\cite{pagels-stokar} with new $\Sigma(p^2)$-dependent vertex we
suggest here that the SM fermion mass splitting in families is
entirely due to the electroweak interactions with peculiar vectorial
$\Sigma(p^2)$-dependent vertices enforced by the WT identities
\cite{pagels-stokar}, \cite{ball-chiu}, \cite{delbourgo},
\cite{benes-hosek}.

The new $\Sigma_f(p^2)$-dependent polar-vector electroweak vertices
are
\begin{widetext}
\begin{eqnarray}
\Gamma_{A}'^{\mu}(p',p) &=& -eQ_i (p' + p)^{\mu}\Sigma'(p',p) \,,
\\
\Gamma_{W}'^{+\mu}(p',p) &=& -\frac{e}{2 \surd 2 \sin\theta_W}(p' + p)^{\mu}\Sigma'(p',p)T^{+} \,,
\\
\Gamma_{Z}'^{\mu}(p',p) &=& -\frac{e}{\sin 2 \theta_W} (p' + p)^{\mu}\Sigma'(p',p)\big(T_{3L}^i - 2 Q_i  \sin^2 \theta_W\big) \,.
\end{eqnarray}
\end{widetext}
Being {\it chirality-changing} they are perfectly suited for
computing different fermion masses in a given family in terms of the
{\it known parameters}. An obvious objection is of course how it can
be that the weakly coupled electroweak interaction describes the
observed {\it enormous} top-bottom quark mass splitting. There
is a hope. The computation of the fermion pole-mass splitting
amounts to solving a complicated algebraic equation (\ref{pole})
which crucially depends upon the functional form of the resulting
$\Sigma_f^i(p^2)$. It of course critically depends upon the functional
form of $\Sigma_f(p^2)$. Another obvious objection is that the observed
ordering of fermion masses in different families is {\it different}
whereas the electroweak interactions are {\it identical} for all
three families: In the first family the $u$ quark is lighter than
the $d$ quark, whereas in other two families the corresponding
ordering is dramatically reversed. In the second family the muon is
heavier than the $s$ quark, whereas in other two families the
corresponding ordering is reversed. There is a hope. The `known'
parameters $m_Z/m_f$ and $m_W/m_f$ which come from QFD and from the axial-vector terms
in the WT identities could account for these properties. \\

It is utmost important that the fermion mass splitting due to
the {\it weakly coupled} electroweak interactions is not
spontaneous: It is not associated with any additional spontaneous
electroweak gauge symmetry breaking. The very formation of both the
composite `would-be' NG bosons and of the composite genuine Higgs
bosons, exhibiting spontaneous breakdown of $SU(3)_f \times SU(2)_L
\times U(1)_Y$ down to $U(1)_{em}$ can solely be due to the strong
coupling, i.e., to QFD.

Important consequence of this reasoning is the neutrino mass
spectrum. As described below the Dirac neutrino mass matrix
$m_f^{\nu}$ results from characteristic contributions from $W$ and
$Z$ exchanges and, because of the Majorana mass matrix $M_{fR}$ of
sterile right-handed neutrinos {\it the neutrino mass spectrum of
the model is given by the famous seesaw mass matrix} \cite{seesaw}.\\

We treat the contributions of the photon, $W$ and $Z$ boson on the
same footing and take the massive propagators in the `soft' form of
a massive vector boson coupled to a conserved current:
\begin{equation}
D^{\mu \nu}_{W,Z}(q) = -\I\frac{g^{\mu \nu}-q^\mu q^\nu/q^2}{q^2 - m_{W,Z}^2}
\,,
\end{equation}
i.e., in parallel with the massless photon propagator in Landau
(transverse) gauge. Namely this form is obtained when summing the
contributions of the massless `would-be' NG poles in the $W,Z$ boson
polarisation tensors in order to obtain the massive $W,Z$ boson
poles \cite{eb}.
It also enables immediate control with photon contribution by setting $m_{W,Z}$ to zero.\\

The photon $A$ and the $Z, W$-boson corrections to the Dirac fermion
self-energies $\delta_{A,Z,W}^i \Sigma_f(p^2)$, $i=\nu, l, u, d$ are
characterized by the strengths $-e^2Q_i^2$ and $-e^2P_{iZ,W}^2$,
respectively, where (for $\sin^2 \theta_W \sim 0.23$)
\begin{eqnarray*}
Q^2_{\nu}&=& 0\,,\phantom{h}     P^2_{\nu Z}=1/4\sin^2 2\theta_W \cong 0.35 \,,
\\
Q^2_l &=& 1\,, \phantom{h}      P^2_{l Z} = (1-4\sin^2 \theta_W)^2/4\sin^2 2\theta_W \cong 0.00 \,,
\\
Q^2_u &=& \tfrac{4}{9}\,, \phantom{h}     P^2_{u Z} = (1-\tfrac{8}{3}\sin^2 \theta_W)^2/4\sin^2 2\theta_W \cong 0.05 \,,
\\
Q^2_d &=& \tfrac{1}{9}\,,\phantom{h}     P^2_{d Z} = (1-\tfrac{4}{3}\sin^2 \theta_W)^2/4\sin^2 2\theta_W \cong 0.17 \,.
\end{eqnarray*}

The $W$-boson correction to the Dirac self-energies $\delta_W
\Sigma_f(p^2)$ is characterized by the strength which does not
depend upon the fermion species distinguishing index $i$:
$-e^2P^2_W$ where $P^2_W=1/8\sin^2\theta_W \cong 0.54$

For the SM fermion $i=\nu, l, u, d$ in family $f$ the contributions
$\delta^{(i)}_{A,W,Z} \Sigma_f(p^2)$ to the common
$\Sigma_f(p^2)=m_f^2/p$ due to the new $\Sigma_f(p^2)$-dependent vertices are given in the lowest order by the Feynman diagrams of Fig.~1.

\begin{figure}[t]
\begin{center}
\includegraphics[width=1\linewidth]{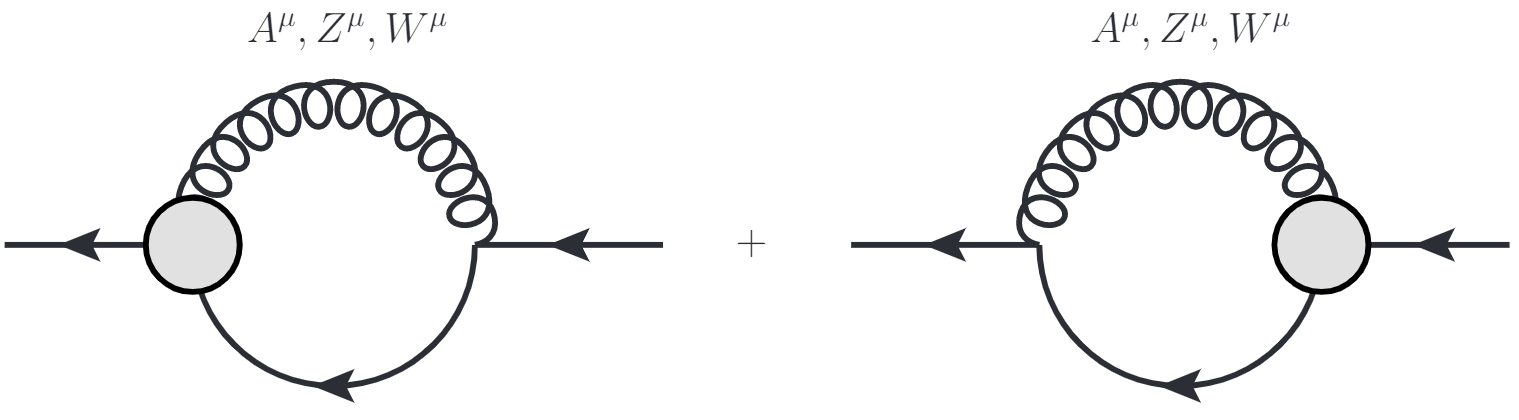}
\end{center}
\caption{The lowest-order contributions of new $\Sigma$-dependent vertices to the fermion mass splitting in $f$.}
\label{fig_vertex3}
\end{figure}

\begin{widetext}
\begin{eqnarray}
\delta^i_A \Sigma_f(p^2)&=& -e^2 Q_i^2\int
\frac{\d^4k}{(2\pi)^4}\frac{1}{(p-k)^4}\frac{\Sigma_f(p^2)-\Sigma_f(k^2)}{k^2+\Sigma_f(k^2)}\frac{p^2k^2-(p.k)^2}{p^2-k^2} \,,
\\
\delta^i_{Z,W} \Sigma_f(p^2) &=& -e^2 P_{iZ,W}^2\int
\frac{\d^4k}{(2\pi)^4}\frac{1}{(p-k)^2}\frac{1}{(p-k)^2 +
m_{Z,W}^2}\frac{\Sigma_f(p^2)-\Sigma_f(k^2)}{k^2+\Sigma_f(k^2)}\frac{p^2k^2-(p.k)^2}{p^2-k^2} \,.
\end{eqnarray}
\end{widetext}
It is rather remarkable that these integrals can be computed
exactly. In order to obtain a comprehensible, illustrative fermion
mass formula we make some elementary simplifying approximations. All
necessary details are given in the Appendix. The result is the {\it Euclidean} $\Sigma_f^i(p^2)$:
\begin{equation}
\Sigma_f^i(p^2)=\frac{m_f^2}{p}+\delta^{i(IR)}_A
\Sigma_f(p^2)+\delta^{i(IR)}_{Z} \Sigma_f(p^2)+\delta^{(IR)}_{W}
\Sigma_f(p^2)
\,.
\label{sigmaif}
\end{equation}

The fermion masses $m_f^i$ are given as the poles of the full
fermion propagators with the momentum-dependent chirality-changing
self-energies $\Sigma_f^i(p^2)$ in the {\it Minkowski} space. For this reason we replace in
(\ref{sigmaif}) the Euclidean $p=\surd p^2$ by $p=\surd -p^2$ where the Minkowski $p^2$
can be both  positive and negative. The resulting pole equation to be solved for $p^2=m_f^{i2}$
with fixed $i$, $f$ has the form
\begin{equation}
p^2=\Sigma_f^{i+}(p^2) \, \Sigma_f^i(p^2) \,.
\label{pole}
\end{equation}
Here
\begin{equation}
\Sigma_f^i(p^2)=-\I\frac{m_f^2}{\surd p^2} A_f^i(p^2) + \I\surd p^2 B_f^i(p^2) \,.
\end{equation}
The real dimensionless functions $A$, $B$ are
\begin{widetext}
\begin{eqnarray}
A_f^i(p^2) &=& 
1 + \frac{3\alpha}{8\pi} Q_i^2 \bigg[1-\frac{m_f^2}{p^2} \arctg\frac{p^2}{m_f^2}\bigg]
\nonumber \\ && {}
\phantom{1}
+ \frac{3\alpha}{8\pi} P_{iZ}^2 \bigg\{ -\frac{1}{2} \bigg(1-\frac{m_f^2}{m_Z^2}\bigg) \arctg\frac{p^2}{m_f^2}
- \frac{m_f^2}{p^2} \bigg[\arctg\frac{p^2}{m_f^2} + \frac{3m_Z^4+m_f^4}{6m_f^2 m_Z^2} \ln\bigg(1+\frac{p^4}{m_f^4}\bigg)\bigg] \bigg\}
\nonumber \\ && {}
\phantom{1}
+ \frac{3\alpha}{8\pi} P_{W}^2 \bigg\{ -\frac{1}{2}\bigg(1-\frac{m_f^2}{m_W^2}\bigg) \arctg\frac{p^2}{m_f^2}
- \frac{m_f^2}{p^2}\bigg[\arctg\frac{p^2}{m_f^2} + \frac{3m_W^4+m_f^4}{6m_f^2 m_W^2} \ln\bigg(1+\frac{p^4}{m_f^4}\bigg)\bigg] \bigg\} \,,
\\
B_f^i(p^2) &=& \frac{3\alpha}{32\pi} \bigg(P_{iZ}^2\frac{m_f^2}{m_Z^2} + P_W^2\frac{m_f^2}{m_W^2}\bigg) \bigg[\frac{5}{3}-\ln\bigg(1+\frac{p^4}{m_f^4}\bigg)\bigg] \,.
\end{eqnarray}
\end{widetext}

Our ignorance about $M_{fR}$ does not allow to specify the neutrino propagator in which the neutrino mass is given by the seesaw mass formula.
As such a propagator is needed in computing the $W$ contribution to the charged lepton masses, we have to restrict ourselves at present merely
to the computing of the quark mass splitting.

For an illustration we present the approximate explicit solution of the pole equation (\ref{pole}) assuming $m_f^2/m_f^{i2}\ll 1$. This amounts to solving
the simplified equation $B_f^i(m_f^{i2})=\pm 1$. It is interesting that in this approximation the dependence upon $\alpha$ disappears.
The consistency with the assumption demands the sign plus, and the result is
\begin{equation}
m_f^i = m_f e^{5/12} \exp
\Bigg\{
\frac{40\pi \sum m_f^2}{3m_f^2 \big[\big(T_{3L}^i-2Q_i \sin^2 \theta_W\big)^2 + \tfrac{1}{2}\big]}
\Bigg\} \,.
\end{equation}
Although aesthetically appealing this mass formula does not approximate the quark world. This, however, suggests that the all-important
infrared part of $\Sigma_f^i(p^2)$ could be guessed.

In general, with the Euclidean $\Sigma_f(p^2)=m_f^2/p$ resulting from a crude separable approximation
the hopes remain unfulfilled: the numerical solutions of the equation (\ref{pole}) exhibit only the small,
nonrealistic fermion mass splitting.

\section{Conclusion}

Replacement of the weakly coupled Higgs sector of the SM with many parameters by the
strong-coupling gauge QFD with one parameter is an immodest
challenge. We believe it is in the spirit of the old good traditions
of theoretical physics. As a quantum field theory it makes sense if and only if
the fermion sector of the Standard model is uniquely extended, for
purely theoretical reason of anomaly freedom, by three right-handed
neutrinos. {\bf Spontaneously, i.e., in solutions} the matrix SD
equation of QFD, {\bf so far in a separable approximation},
generates from one scale $\Lambda$ {\it in two different channels
two different sets of solutions with uniquely fixed and vastly
different fermion masses}: First, three Majorana masses $M_{fR}\sim
\Lambda$ of $\nu^f_R$ . Second, three exponentially small Dirac
masses $m_f \sim \exp{(-1/\alpha_f)}\Lambda $ ($\alpha_f$ are the
effective dimensionless couplings of separable approximation)
degenerate for all SM fermions in $f$. The soft Majorana masses
$\Sigma_{fR}(p^2) = M^2_{fR}/p$ spontaneously emerge despite the
fact that the hard ones are strictly prohibited by the hidden chiral
symmetry of the $\nu^f_R$ sector of QFD; the soft SM Dirac masses
$\Sigma_f(p^2)=m^2_f/p$ spontaneously emerge despite the fact that
the hard ones are strictly prohibited by the chiral SM.\\

We know no way of knowing whether this highly desirable property is
inherent to QFD at strong coupling or not. In any case the same
property $M_{fR}\gg m_f$ was obtained by Yanagida \cite{yanagida}
easily in a very useful model with fermion content identical to
ours, but with the rich Higgs sector with elementary Higgs fields in
representations dictated by symmetry of allowed Yukawa couplings.
Not surprisingly our fermion-composite Higgs-type operators have the
same quantum numbers.

If our main conjecture is warranted {\bf there is no fundamental
electroweak mass scale}. The only genuine scale in the game is the
huge QFD scale $\Lambda$. Although theoretically arbitrary the
phenomenology of the emerging picture (e.g. seesaw or FCNC) forces $\Lambda$
to be fixed by one experimental datum as huge: The Dirac masses
$m_f$ of the SM fermions come out spontaneously exponentially small
with respect to $\Lambda$. Goldstone theorem implies the $W,Z$
masses of the order of $\sum_f m_f$, the {\it induced} or {\it
effective} electroweak mass scale. The Higgs boson is light because
it is intrinsically related with this scale as the genuine partner
of the longitudinal polarization states of massive $W$ and $Z$
bosons. There is no fine tuning. In the resulting `improved'
Standard model the facts `ignored' by the Standard one mentioned in
the Introduction are addressed and bona fide explained:

(1) By seesaw the present model describes its three active neutrinos
as extremely light {\it Majorana} fermions. In comparison with the
general effective field theory prediction of Steven Weinberg
\cite{weinberg1} the present ordinary quantum field theory is
uniquely defined. It simply needs three right-handed neutrinos for
anomaly freedom in much the same way the famous Weinberg's model of
leptons \cite{sm} needs quarks \cite{bouchiat}.

(2) By the same logic the super-heavy right-handed neutrinos should
be phenomenologically equally important as ordinary SM fermions. In
close parallel with QCD nucleons $N \sim \epsilon_{abc} q^a q^b q^c$
which make the luminous matter of the Universe the model offers the
stable heavy-neutrino composites $S \sim \eta_{abc} \nu_R^a \nu_R^b
\nu_R^c$ ($\eta$ is an appropriate $SU(3)$ Clebsch-Gordan
coefficient) strongly coupled by QFD as a natural candidate for the
dark matter of the Universe. It is conceivable that such a
composite-fermion-made dark Universe might even look not dissimilar
to our luminous one. More generally, the heavy sector with the
fundamental scale $\Lambda$, which is absolutely necessary for
electroweak physics at Fermi scale provides the contact with
astro-particle physics and cosmology. First of all, the heavy
Majorana neutrinos are the necessary ingredient in
Fukugida-Yanagida's baryogenesis via leptogenesis
\cite{leptogenesis}. Furthermore, the classically scale-invariant
matter sector with huge scale $\Lambda$ due to the dimensional
transmutation is indispensable for conservative understanding of the
induced quantum gravity suggested by Sakharov \cite{sakharov} and
advocated at present by Donoghue \cite{donoghue}. Also, the heavy
composite higgses $\chi$ can be identified with inflatons
\cite{barenboim}.

(3) QFD apparently computes the fermion masses $m_f$ and $M_{fR}$,
but that is not sufficient: It does not distinguish between
different SM fermion species within one family. Fortunately, there
are the electroweak gauge interactions. We have computed the mass
splitting of each calculable $m_f$ into $m_f^i$, $i=\nu, e, u, d$
unequivocally in terms of the known parameters entering the photon,
$W$ and $Z$ boson loops with $\Sigma_f(p^2)$ dependent vectorial
vertices. The computation is, unfortunately, hampered by serious
uncertainties. First, with the aim of obtaining the explicit fermion
mass formula the perturbative computation was slightly simplified.
Second and more important, the very form of the original SD equation
and the separable approximation to its kernel result from
approximations at strong coupling which are not under theoretical
control. Hence, the fermion mass formula (\ref{pole}) based on the
explicit form $\Sigma_f(p^2)= m_f^2/p$ should be understood merely
as an illustration of the general idea.
Would we know $M_{fR}$ we could predict,
using the seesaw mass formula, the highly needed neutrino mass spectrum. \\

The present approach touches the Higgs paradigm of the origin of
particle masses. In the first step the strongly coupled QFD
spontaneously and self-consistently generates three Majorana masses
$M_{fR}$ of the right-handed neutrinos and the masses of all flavor
gluons, both of order $\Lambda$. It is important that this step is a
spontaneous breakdown of chiral symmetry. At the same time the QFD
spontaneously generates also three Dirac masses $m_f\ll M_{fR}$ of
the SM fermions. By symmetry they are the same for all SM fermion
species in $f$. The Goldstone theorem implies two types of the
SM-fermion-composite `would-be' NG excitations: (1) Six of them
correspond to the additional breakdown of $SU(3)_f$ by $m_{(3)}$ and
$m_{(8)}$ down to $U(1) \times U(1)$. They belong to the composite
flavor octet and disappear from the spectrum as additional
longitudinal components of massive flavor gluons. Hence, the massive
$h_3$ and $h_8$ remain. (2) All $m_f$ contribute to the spontaneous
breakdown of the electroweak $SU(2)_L \times U(1)_Y$ down to
$U(1)_{em}$. Consequently, there are three SM-fermion-composite NG
bosons and one massive $h$, all belonging to the complex composite
doublet.

In the following step the dynamical role of the electroweak
interactions, although weakly coupled, becomes profound: First, the
$\Sigma_f(p^2)$-dependent terms in the {\it axial-vector} parts of
the electroweak WT identities are responsible for the $W$ and $Z$
boson masses. Second, the $\Sigma_f(p^2)$-dependent terms in the
{\it polar-vector} parts of the electroweak WT identities are
suggested to be responsible for the SM fermion mass splitting of $m_f$ into $m_f^i, i=\nu, l, u, d$.
\\
In evaluating the consequences of the model it is important to distinguish
between those which crucially depend upon the functional form of $\Sigma_f(p^2)$ and those
which rely essentially only upon their very existence and their gross features.
We believe that the unfulfilled expectation in computing the fermion mass splitting in $f$
does not disqualify the whole idea.

IF the basic assumption of the spontaneous, strong-coupling generation
of $M_{fR}\gg m_f$ by QFD is justified,
even without the necessity of knowing the details of $\Sigma_f(p^2)$
there are the firm, and currently observable consequences:
They are due to the powerful existence theorem of Goldstone supplemented
with the natural NJL assumption of the existence of the NG symmetry partners:\\

First of all, after July 4, 2012 the model is obliged to contain the
CERN Higgs boson with properties similar, though not necessarily
identical with those of the elementary SM Higgs $H$. We believe our
$h$ is such a boson. We interpret it as an a posteriori confirmation
of the assumption (in fact due to Nambu and Jona-Lasinio) of the
existence of the NG symmetry partners. Derivation of its
fermion-loop-generated couplings with the electroweak gauge bosons
$A, W, Z$ is in progress \cite{benes-hosek}. Similarity with the
tree-level couplings of $H$ with $W$ and $Z$ of the Standard model
proportional to the masses $m_W$ and $m_Z$, respectively is to be
expected due to the sum rules (\ref{mW},\ref{mZ}).\\

Above all, there is the robust prediction of two new composite
electroweakly interacting Higgs bosons $h_3$ and $h_8$ at Fermi
scale. The model stands and falls with these Higgs-like particles:
They are the remnants of the flavor octet \cite{higgs}, the
composite fermion-antifermion scalar strongly bound by QFD and
responsible for the spontaneous emergence of $m_{(3)}$ and
$m_{(8)}$. These scalars are distinguished by the effective
flavor-sensitive Yukawa couplings proportional to the flavor
matrices $\lambda_3$ and $\lambda_8$, respectively. This signals
that not all three families are alike.\\

The rigid predictive electroweak model presented above represents so
far merely a possible framework, or scenario. The most difficult
step is to provide the convincing argument for $M_{fR}\gg m_f$.
Further, it remains to compute the quantum electroweak corrections
to all quantities in which the Dirac fermion masses $m_f$, generated
by QFD, enter degenerate:

(i) The Yukawa couplings of composite Higgses $h, h_3, h_8$ with fermions.\\
(ii) The effective loop-generated couplings of composite Higgses
$h, h_3, h_8$ with the electroweak gauge fields $A, W, Z$.\\
(iii) The electroweak corrections to the Pagels-Stokar mass formula for $m_W, m_Z$.\\
(iv) We believe that the general idea of the electroweak fermion mass splitting
remains alive. How to look for the convincing functional form of $\Sigma_f(p^2)$
is, however, completely outside our imagination.\\

Reliable computation of masses of $h$, $h_3$ and $h_8$ as well as
the computation of other properties of the expected strongly bound
QFD bound states (e.g. the dark matter neutrino composites) which
are not of the NG nature requires the generically strong-coupling tools;
also a formidable task.\\

\begin{acknowledgments}
The work was partly supported by the grant LG 15052 of the
Ministry of Education of the Czech Republic.
\end{acknowledgments}

\appendix

\section{Evaluation of $\delta^i_{A,W,Z} \Sigma_f(p^2)$}

In the Euclidean integrals for $\delta^i_{A,W,Z} \Sigma_f(p^2)$ we
fix without loss of generality the external four-momentum $p$ as
$p=(p,\vec 0)$, integrate over the angles, and for
$\Sigma_f(p^2)=m_f^2/p$ we get
\begin{eqnarray}
\delta^i_A \Sigma_f(p^2) &=& \frac{2\alpha}{\pi^2}Q_i^2 m_f^2
p\int_0^{\infty}I\frac{k^6 \d k}{(p+k)(k^4 + m_f^4)} \,,
\\
\delta^i_{Z,W}\Sigma_f(p^2) &=& \frac{2\alpha}{\pi^2} P_{iZ,W}^2 m_f^2
p \int_0^{\infty}I_{Z,W}\frac{k^6 \d k}{(p+k)(k^4 + m_f^4)} \,,
\nonumber \\ &&
\end{eqnarray}
where
\begin{widetext}
\begin{eqnarray}
I(p,k) &=& \int_0^{\pi}\frac{\sin^4\Theta \, 
\d\Theta}{(p^2+k^2-2pk \cos\Theta)^2}
=
\frac{3\pi}{16p^4k^4}\Big[p^4+k^4-\big(p^2+k^2\big)\big|p^2-k^2\big| \Big] \,,
\\
I_{Z,W}(p,k) &=& \int_0^{\pi}\frac{\sin^4\Theta \, 
\d\Theta}{(p^2+k^2-2pk \cos\Theta)(p^2+k^2+m_{Z,W}^2-2pk \cos\Theta)}
\\ &=&
\frac{\pi}{16p^4k^4}
\bigg\{ 
3 \bigg[p^4+k^4+\big(p^2+k^2\big)m_{Z,W}^2+\frac{1}{3}m_{Z,W}^4
\nonumber \\ && {}
\hspace{13mm}
+ \frac{1}{m_{Z,W}^2} \bigg[
\big(p^2-k^2\big)^2\big|p^2-k^2\big|-\big((p-k)^2+m_{Z,W}^2\big)^{3/2} \big((p+k)^2+m_{Z,W}^2\big)^{3/2}
s\bigg]
\bigg\}
\,.
\label{izw}
\end{eqnarray}
\end{widetext}
The absolute value $|p-k|$ in both $I(p,k)$ and $I_{Z,W}(p,k)$
naturally leads to splitting the integrals into: (a) $\int_0^p$
which we call infrared (IR) and (b) $\int_p^{\infty}$ which we call
ultraviolet (UV). We consider $p\gg m_W, m_Z$.

In the integral for $\delta^i_A \Sigma_f(p^2)$ we merely replace the
term $(p+k)$ in the integrand by $k$ and obtain
\begin{equation}
\delta^i_A \Sigma_f(p^2) = \delta^{i(IR)}_A
\Sigma_f(p^2)+\delta^{i(UV)}_A \Sigma_f(p^2) \,,
\end{equation}
where
\begin{eqnarray}
\delta^{i(IR)}_A
\Sigma_f(p^2) &=& \frac{3\alpha}{8\pi}Q_i^2\bigg(\frac{m_f^2}{p}-\frac{m_f^4}{p^3}\arctg\frac{p^2}{m_f^2}\bigg) \,,
\\
\delta^{i(UV)}_A \Sigma_f(p^2) &=& \frac{3\alpha}{8\pi}Q_i^2 p \bigg(\frac{\pi}{2}-\arctg\frac{p^2}{m_f^2}\bigg)\,.
\end{eqnarray}
In the integral for $\delta^i_{Z,W}\Sigma_f(p^2)$ we make similar
simplifications: (i) replace the term $(p+k)$ in the integrand by
$k$; (ii) replace the term
$((p-k)^2+m_{Z,W}^2)^{3/2}((p+k)^2+m_{Z,W}^2)^{3/2}$ in
$I_{Z,W}(p,k)$ by $(p^2+m_{Z,W}^2)^3$; (iii) in (UV) we neglect the
gauge boson masses:
\begin{equation}
\delta^i_{Z,W} \Sigma_f(p^2)=\delta^{i(IR)}_{Z,W}
\Sigma_f(p^2)+\delta^{i(UV)}_{Z,W} \Sigma_f(p^2) \,,
\end{equation}
where
\begin{widetext}
\begin{eqnarray}
\delta^{i(IR)}_{Z,W}
\Sigma_f(p^2) &=& \frac{3\alpha}{8\pi} P_{iZ,W}^2 \bigg\{
\frac{1}{2}\frac{m_f^2}{p}\bigg(1-\frac{m_f^2}{m_{Z,W}^2}\bigg)\arctg\frac{p^2}{m_f^2} 
+ \frac{m_f^4}{p^3}\bigg[-\arctg\frac{p^2}{m_f^2} + \frac{1}{2}\bigg(\frac{m_{Z,W}^2}{m_f^2}+\frac{m_f^2}{3m_{Z,W}^2}\bigg) \ln\bigg(1+\frac{p^4}{m_f^4}\bigg)\bigg]
\nonumber \\ && {}
\hspace{10mm}
+ \frac{1}{4} p \frac{m_f^2}{m_{Z,W}^2} \bigg[\frac{5}{3} - \ln\bigg(1+\frac{p^4}{m_f^4}\bigg)\bigg]
\bigg\} \,,
\\
\delta^{i(UV)}_{Z,W} \Sigma_f(p^2) &=& \frac{3\alpha}{8\pi} P_{iZ,W}^2 p \bigg(\frac{\pi}{2}-\arctg\frac{p^2}{m_f^2} \bigg) \,.
\end{eqnarray}
\end{widetext}
In accordance with approximations used in solving the SD equation
for $\Sigma_f(p^2)$ the ultraviolet pieces of $\Sigma_f^i(p^2)$ can
be ignored. Consequently, the resulting $\Sigma_f^i(p^2)$ which
defines the SM fermion mass splitting is given by the formula
(\ref{sigmaif}).


\begin{thebibliography}
\bibitem{argument} The fermion mass terms themselves are super-renormalizable
but, being non-invariant they break down the gauge symmetry in
loops. This implies the necessity of introducing the non-invariant
gauge boson mass counter-terms, i.e., the unitarity-violating hard
gauge boson masses.
\end{thebibliography}
\end{document}